\documentclass[a4paper,11pt]{article}

\usepackage{contribution}
\usepackage{xspace}

\newcommand{\weblink}[2][]{%
    \ifthenelse{\equal{#1}{}}%
    {\textnormal{\url{#2}}}%
    {\textnormal{\href{#2}{#1}}}%
}

\def\beq{\begin{equation}}
\def\eeq#1{\label{#1}\end{equation}}
\def\eeqn{\end{equation}}

\def\beqa{\begin{eqnarray}}
\def\eeqa#1{\label{#1}\end{eqnarray}}
\def\eeqan{\end{eqnarray}}

\let\bar=\overbar

\def\Dslash{\not{\hbox{\kern-4pt $D$}}}
\def\dslash{\not{\hbox{\kern-2pt $\del$}}}

\def\msb{{\bar{\ssstyle M \kern -1pt S}}}

\newcommand{\contribution}[7][]{%
  \clearpage
  \thispagestyle{plain}
  \ifthenelse{\equal{#1}{}}
  {\hypersetup{pdftitle={#2}}}
  {\hypersetup{pdftitle={#1}}}
  \hypersetup{pdfauthor={{#3} {#4}}}
  {\centering\normalfont\LARGE\bfseries\sffamily #2 \par\nobreak}
  \lhead{}
  \chead{%
    \textit{\footnotesize XIV International Conference on Hadron Spectroscopy
      (\weblink[\textit{hadron2011}]{http://www.hadron2011.de}), 13-17 June 2011, Munich, Germany}%
  }
  \rhead{}
  \bigskip
  \begin{center}
    {#3} {#4}\ifthenelse{\equal{#6}{}}{}{\footnote{\weblink[#6]{mailto:#6}}}
    \ifthenelse{\equal{#7}{}}{}{#7} \\
    \textit{#5}
  \end{center}
  \bigskip
}

\renewcommand{\abstract}[1]{%
  \begin{center}
    \begin{minipage}{0.85\textwidth}
      \begin{footnotesize}
        #1
      \end{footnotesize}
    \end{minipage}
  \end{center}
  \bigskip
}

\begin{document}

%
%
%
%
%
{  

\newcommand{\Jpsi}{\ensuremath{J\!/\!\psi}\xspace}
\newcommand{\Lc}{\ensuremath{\Lambda_c^+}\xspace}
\newcommand{\LcS}{\ensuremath{\Lambda_c^{*+}}\xspace}
\newcommand{\Lcs}{\ensuremath{\Lambda_c^{(*)+}}\xspace}
\newcommand{\Lcl}{\ensuremath{\Lambda_c^{*+}(2595)}\xspace}
\newcommand{\Sc}{\ensuremath{\Sigma_c}\xspace}
\newcommand{\ScS}{\ensuremath{\Sigma_c^*}\xspace}
\newcommand{\Scs}{\ensuremath{\Sigma_c^{(*)}}\xspace}
\newcommand{\Xic}{\ensuremath{\Xi_c}\xspace}
\newcommand{\Xicp}{\ensuremath{\Xi_c^+}\xspace}
\newcommand{\Xicz}{\ensuremath{\Xi_c^0}\xspace}
\newcommand{\Lb}{\ensuremath{\Lambda_b}\xspace}
\newcommand{\Sb}{\ensuremath{\Sigma_b}\xspace}
\newcommand{\Sbp}{\ensuremath{\Sigma_b^+}\xspace}
\newcommand{\Sbm}{\ensuremath{\Sigma_b^-}\xspace}
\newcommand{\SbS}{\ensuremath{\Sigma_b^*}\xspace}
\newcommand{\SbSp}{\ensuremath{\Sigma_b^{*+}}\xspace}
\newcommand{\SbSm}{\ensuremath{\Sigma_b^{*-}}\xspace}
\newcommand{\Sbs}{\ensuremath{\Sigma_b^{(*)}}\xspace}
\newcommand{\Sbsp}{\ensuremath{\Sigma_b^{(*)+}}\xspace}
\newcommand{\Sbsm}{\ensuremath{\Sigma_b^{(*)-}}\xspace}
\newcommand{\Xib}{\ensuremath{\Xi_b}\xspace}
\newcommand{\Xibm}{\ensuremath{\Xi_b^-}\xspace}
\newcommand{\Xibz}{\ensuremath{\Xi_b^0}\xspace}
\newcommand{\Ob}{\ensuremath{\Omega_b^-}\xspace}
\newcommand{\LbJpsiL}{\ensuremath{\Lb \rightarrow \Jpsi\Lambda}\xspace} 
%

\contribution[Heavy Flavor Baryons]  
{Heavy Flavor Baryons \\ at the Tevatron}  
{Thomas}{Kuhr}  
{Karlsruhe Institute of Technology \\
  Institut f\"ur Experimentelle Kernphysik \\
  Wolfgang-Gaede-Str. 1 \\
  D-76131 Karlsruhe, GERMANY}  
{Thomas.Kuhr@kit.edu}  
{on behalf of the CDF and D0 Collaborations}  
%

\abstract{%
The Tevatron experiments CDF and D0 have filled many empty spots in the spectrum of heavy baryons over the last few years.
The most recent results are described in this article:
The first direct observation of the \Xibz, improved measurements of \Sb properties, a new measurement of the \LbJpsiL branching ratio, and a high-statistics study of charm baryons.
}
%

\section{Introduction}

The \Lb baryon, with quark content $\left| udb \right>$, is known since about 20 years.
But only recently considerable progress could be made on the other components of the $b$-baryon family by the Tevatron experiments CDF and D0.
The charged \Sb baryons ($\left| uub \right>$ and $\left| ddb \right>$) were observed in the decay to $\Lb\pi^\pm$ by CDF in 2007~\cite{Aaltonen:2007rw}.
In the same year the charged \Xib state ($\left| dsb \right>$) was directly observed by D0~\cite{Abazov:2007ub} and CDF~\cite{Aaltonen:2007un} in the decay to $\Jpsi \Xi^-$.
Until then only indirect observations of the \Xib via an excess in $\Xi^- \ell^-$ events were reported by the ALEPH~\cite{Buskulic:1996sm} and DELPHI~\cite{Abreu:1995kt} experiments.
In 2008 the \Ob ($\left| ssb \right>$) was discovered by D0 in the decay to $\Jpsi \Omega^-$~\cite{Abazov:2008qm}.
CDF observed the \Ob briefly afterwards~\cite{Aaltonen:2009ny}, but measured a mass that is incompatible with the value quoted by D0.
In the last year the Tevatron experiments continued to improve the knowledge about heavy baryons by further measurements, which are presented in this article.

The progress on the field of heavy baryons has been possible because of the large dataset delivered by the Tevatron $p\bar{p}$ collider running at a center of mass energy of $\sqrt{s}=1.96$ TeV.
During the Run II period, both experiments, CDF and D0, collected about 10 fb$^{-1}$ of data.
The Tevatron is well suited for heavy baryon studies, e.g. compared to B factories, because all kinds of heavy hadrons are produced with significant cross section.
On the other hand the high combinatorial background and the huge inelastic cross section are a challenge.
To be able to record the interesting events with heavy flavor hadrons, highly selective and efficient triggers are essential~\cite{Triggers}.
Both experiments can trigger on dimuon pairs so that heavy baryon decays to \Jpsi mesons can be studied.
While D0 has an efficient single muon trigger, CDF is able to trigger on hadronic decays of heavy hadrons identified by tracks displaced from the primary vertex.

\section{\Xibz Observation}

The \Xibm was directly observed by D0 and CDF in the decay to $\Jpsi \Xi^-$.
The events were triggered by the dimuon decay of the \Jpsi and the $\Xi^-$ was reconstructed via $\Xi^- \rightarrow \Lambda \pi^-$.
Since the corresponding decay of the neutral isospin partner, \Xibz ($\left| usb \right>$) $\rightarrow \Jpsi \Xi^0$ with $\Xi^0 \rightarrow \Lambda \pi^0$, involves a neutral pion which cannot be detected efficiently, a search for the \Xibz requires a trigger on a hadronic decay.
CDF reported the first observation of the \Xibz~\cite{Aaltonen:2011wd} in the decay chain $\Xibz \rightarrow \Xicp\pi^-$, $\Xicp \rightarrow \Xi^- \pi^+ \pi^+$, and $\Xi^-\rightarrow \Lambda \pi^-$ briefly after the Hadron2011 conference and therefore this result is included in this article although it was not shown at the conference.
In the same analysis of 4.2 fb$^{-1}$ of data, CDF also confirms the observation of the \Xibm baryon, reconstructed for the first time in its hadronic final state $\Xicz\pi^-$ with $\Xicz \rightarrow \Xi^-\pi^+$.

\begin{figure}[htb]
\begin{center}
\includegraphics[width=0.85\textwidth]{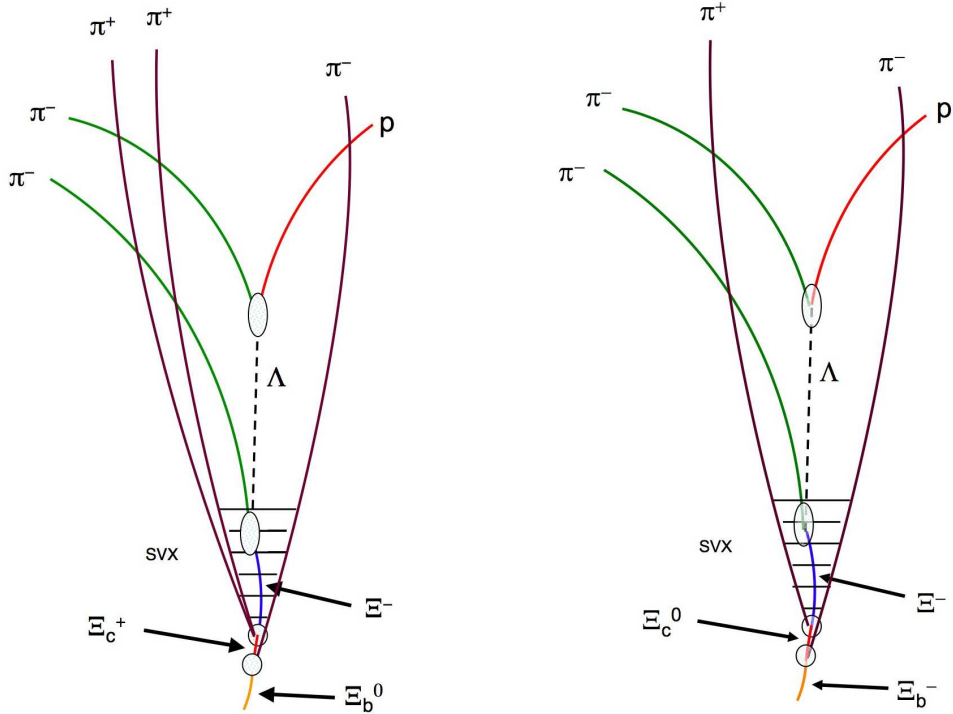}
\caption{Illustration of the \Xibz (left) and \Xibm (right) decay topology.}
\label{fig:XibCartoon}
\end{center}
\end{figure}

The reconstruction starts with the selection of $\Lambda$ candidates from $p\pi^-$ pairs.
$\Xi^-$ candidates are constructed from $\Lambda\pi^-$ pairs which are then combined with one or two $\pi^+$ tracks to form \Xicz or \Xicp candidates, respectively.
The addition of a further $\pi^-$ track to the $\Xi_c$ baryons yields the \Xib candidates.
A schematic illustration of the reconstructed decay chain is shown in Fig.~\ref{fig:XibCartoon}.

The long lifetime of hyperons and charm baryons is exploited by requirements on the reconstructed flight lengths, decay times, or impact parameters.
The $\Xi^-$ candidate is required to be identified by hits in the silicon vertex detector (SVX) which significantly reduces background as illustrated in Fig.~\ref{fig:Xic}.
A simultaneous vertex fit of all tracks with mass constraints for the $\Lambda$, $\Xi^-$, and \Xic is performed to improve the momentum resolution of the \Xib.
The reconstructed invariant mass spectra of \Xib candidates are shown in Fig.~\ref{fig:Xib}.

\begin{figure}[htb]
\begin{center}
\includegraphics[width=0.48\textwidth]{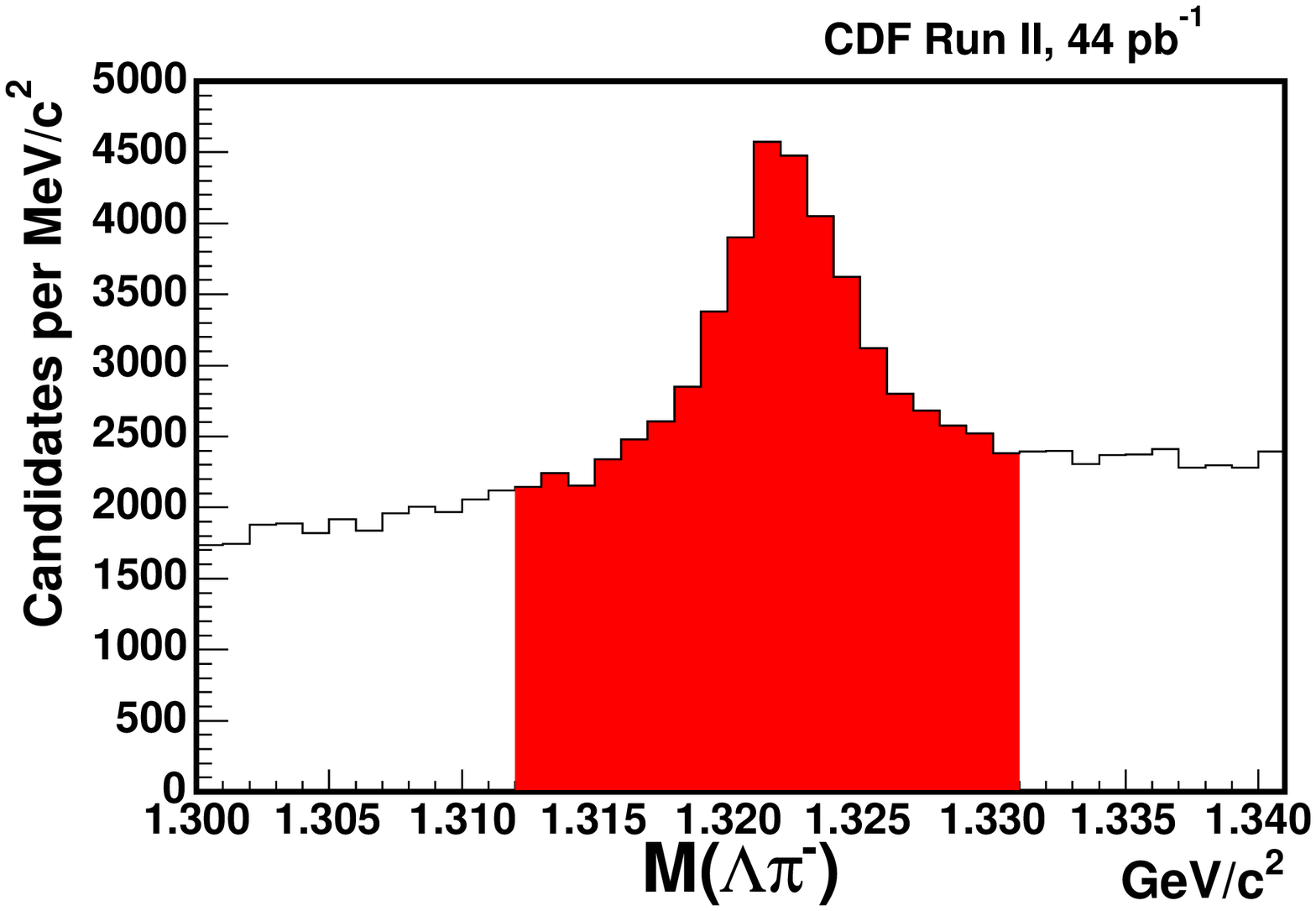}
\includegraphics[width=0.48\textwidth]{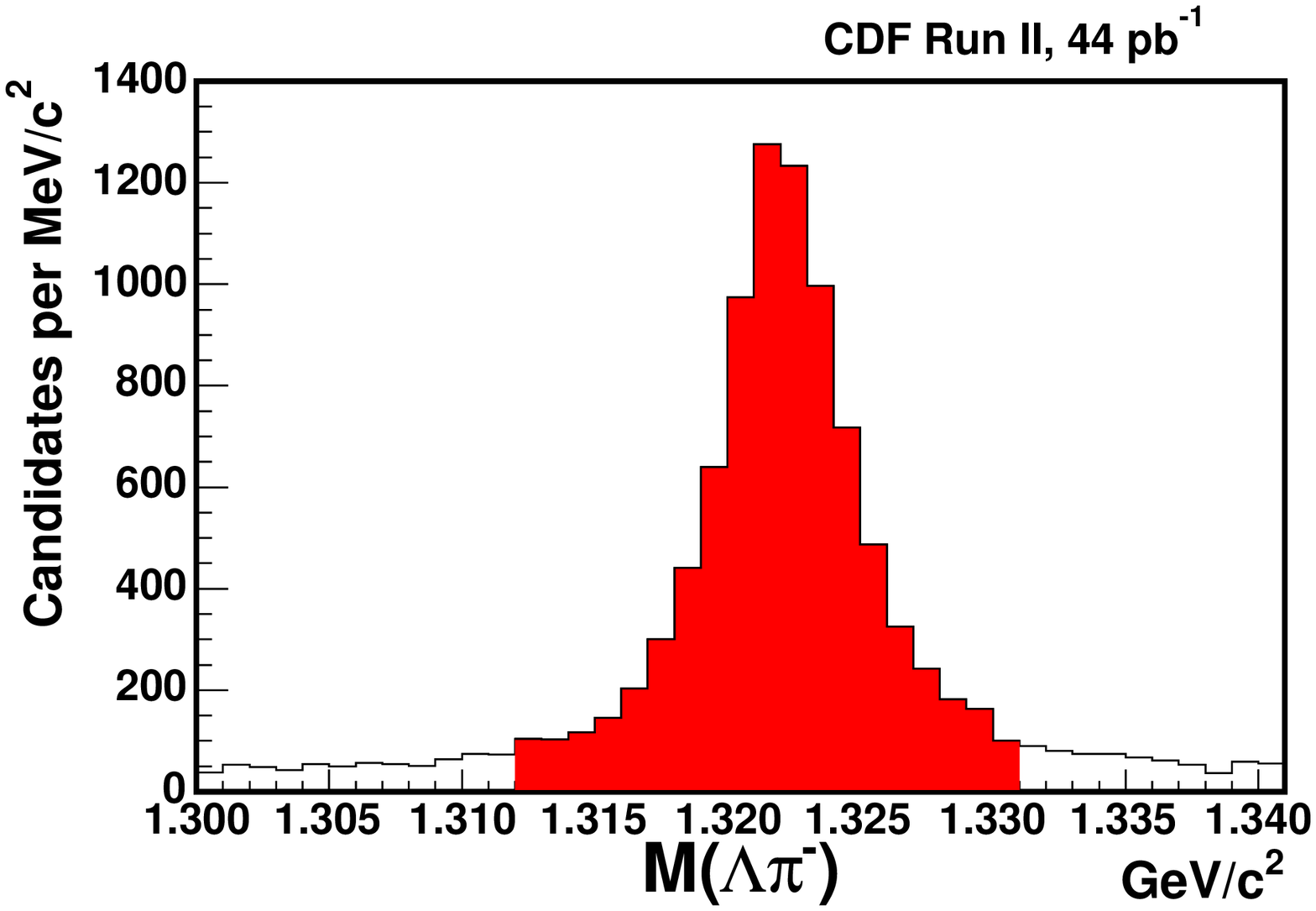}
\caption{The invariant mass distributions of $\Xi^-$ candidates without (left) and with (right) the requirement of SVX hits.}
\label{fig:Xic}
\end{center}
\end{figure}

\begin{figure}[htb]
\begin{center}
\includegraphics[width=0.8\textwidth]{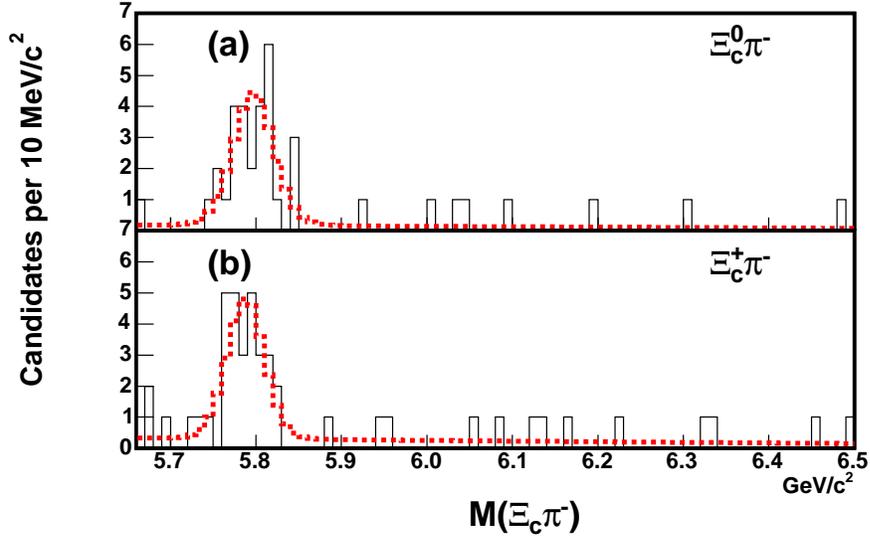}
\caption{The $\Xicz\pi^-$ (a) and $\Xicp\pi^-$ (b) mass distributions with fit projection as dashed line.}
\label{fig:Xib}
\end{center}
\end{figure}

A clear signal is evident for the \Xibm and \Xibz.
To determine the significance and to measure the mass of the states, an unbinned likelihood fit is performed.
The signal is described by a Gaussian and the background by a linear function.
The width of the Gaussian is given by the reconstructed mass resolution of each candidate multiplied with a common scale factor which is fitted to a value consistent with 1.
The significance is determined from a likelihood ratio to be at least 6.8$\sigma$.
The yields are 25.8$^{+5.5}_{-5.2}$ \Xibm and 25.3$^{+5.6}_{-5.4}$ \Xibz baryons.
Their masses are measured to be $m(\Xibm) = (5796.7 \pm 5.1 \pm 1.4)$~GeV$/c^2$ and
$m(\Xibz) = (5787.8 \pm 5.0 \pm 1.3)$~GeV$/c^2$.
The systematic uncertainties are given by the absolute mass scale, the mass resolution scale,
and the world average \Xic masses.
The \Xibm mass is well consistent with the one measured in the $\Jpsi \Xi^-$ decay channel~\cite{Aaltonen:2007un}.

\section{\Sbs Masses and Widths}

\Sbs baryons form an isospin triplet and have a flavor symmetric light di-quark with spin 1.
This couples with the heavy quark spin to two possible spin states, $J^P=\frac{1}{2}^+$ and $J^P=\frac{3}{2}^+$, referred to as \Sb and \SbS, respectively.
The \Sbs baryons decay strongly to the \Lb ground state via the emission of a pion.
Figure~\ref{fig:bBaryonSpectrum} illustrates the spectrum of baryons consisting of $u$, $d$, and one $b$ quark.

\begin{figure}[htb]
\begin{center}
\includegraphics[width=0.8\textwidth]{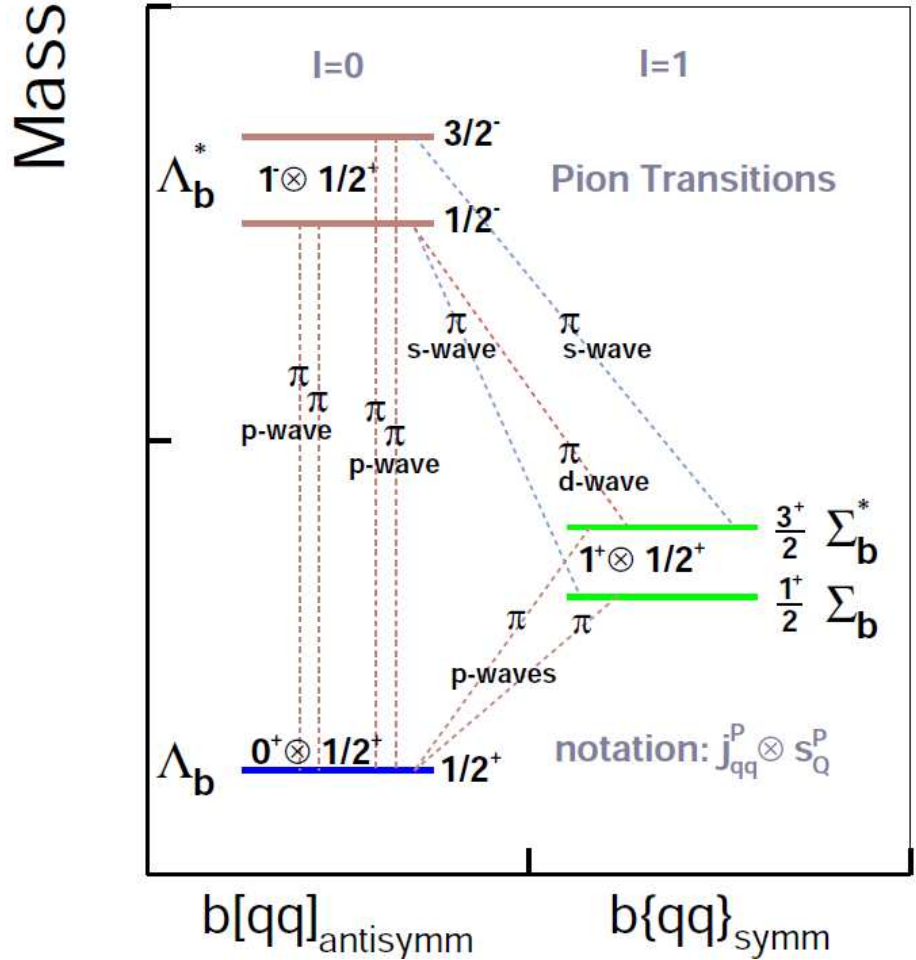}
\caption{Spectrum and decays of $b$ baryons with no strangeness.}
\label{fig:bBaryonSpectrum}
\end{center}
\end{figure}

The decays of the charged \Sbs states were first observed by CDF in 2007~\cite{Aaltonen:2007rw}.
In a data sample of 1.1~fb$^{-1}$, the significance of each of the four states was about 3$\sigma$.
Assuming the same hyperfine splitting between \SbS and \Sb for both charged states, the masses were measured.
Now CDF presented an updated analysis of 6~fb$^{-1}$ of data with improved significances, unconstrained mass measurements, and first measurements of the \Sbs natural widths~\cite{cdf10286}.

The \Lb baryons from the $\Sbs \rightarrow \Lb\pi^\pm$ decay are reconstructed in the decay to $\Lc\pi^-$ with $\Lc \rightarrow pK^-\pi^+$.
The tracks of the final state particles from the \Lc and \Lb decay are usually displaced from the primary vertex so that these decays are selected by the hadronic trigger.
The selection requirements on lifetime and kinematic variables are optimized on the significance of the \Lb signal.
The reconstructed \Lb invariant mass distribution is shown in Fig.~\ref{fig:Lambdab}.
The selected sample contains $16\,000$ \Lb baryons with a signal to background ratio of about $1.8$.
Thus the main background to the \Sbs signals in $\Lb\pi^\pm$ combinations are real \Lb with a random pion.

\begin{figure}[htb]
\begin{center}
\includegraphics[width=0.8\textwidth]{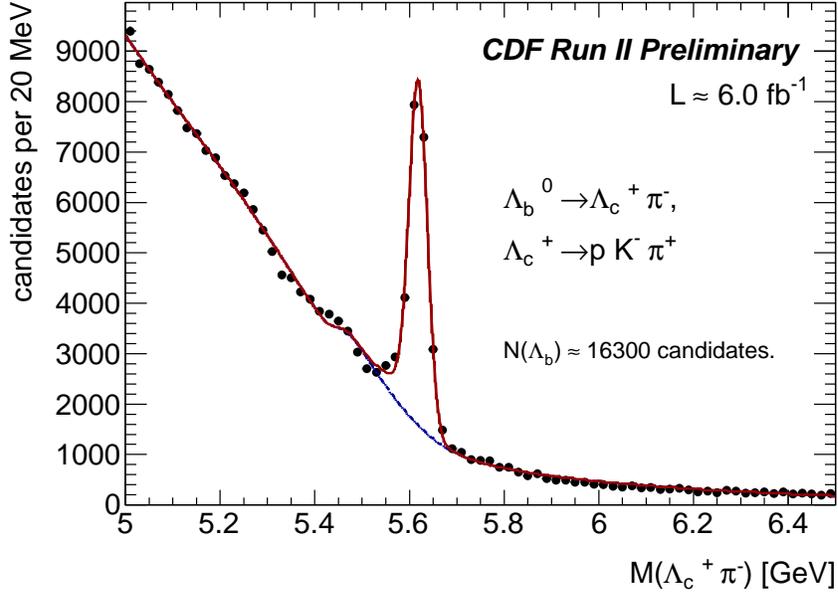}
\caption{Invariant mass distribution of \Lb candidates.}
\label{fig:Lambdab}
\end{center}
\end{figure}

To measure masses and width, the $Q$ value distribution is fitted, where $Q$ is the difference between the reconstructed $\Lb\pi^\pm$ mass and the sum of \Lb and pion masses.
The background is described by an empirical function consisting of a second order polynomial times a square root function to describe the threshold behavior.
Compared to the previous analysis, the background description does not rely on simulation any more.
Each of the four signal peaks is parametrized by a non-relativistic Breit-Wigner.
To account for the $p$ wave decay, a variable width is used which scales with $p_\pi^3$ where $p_\pi$ is the pion momentum in the \Sbs rest frame.
The natural line shape is convolved with a double Gaussian resolution function whose parameters are determined from simulation.
Projections of the fit are shown in Fig.~\ref{fig:Sigmab}.

\begin{figure}[htb]
\begin{center}
\includegraphics[width=0.49\textwidth]{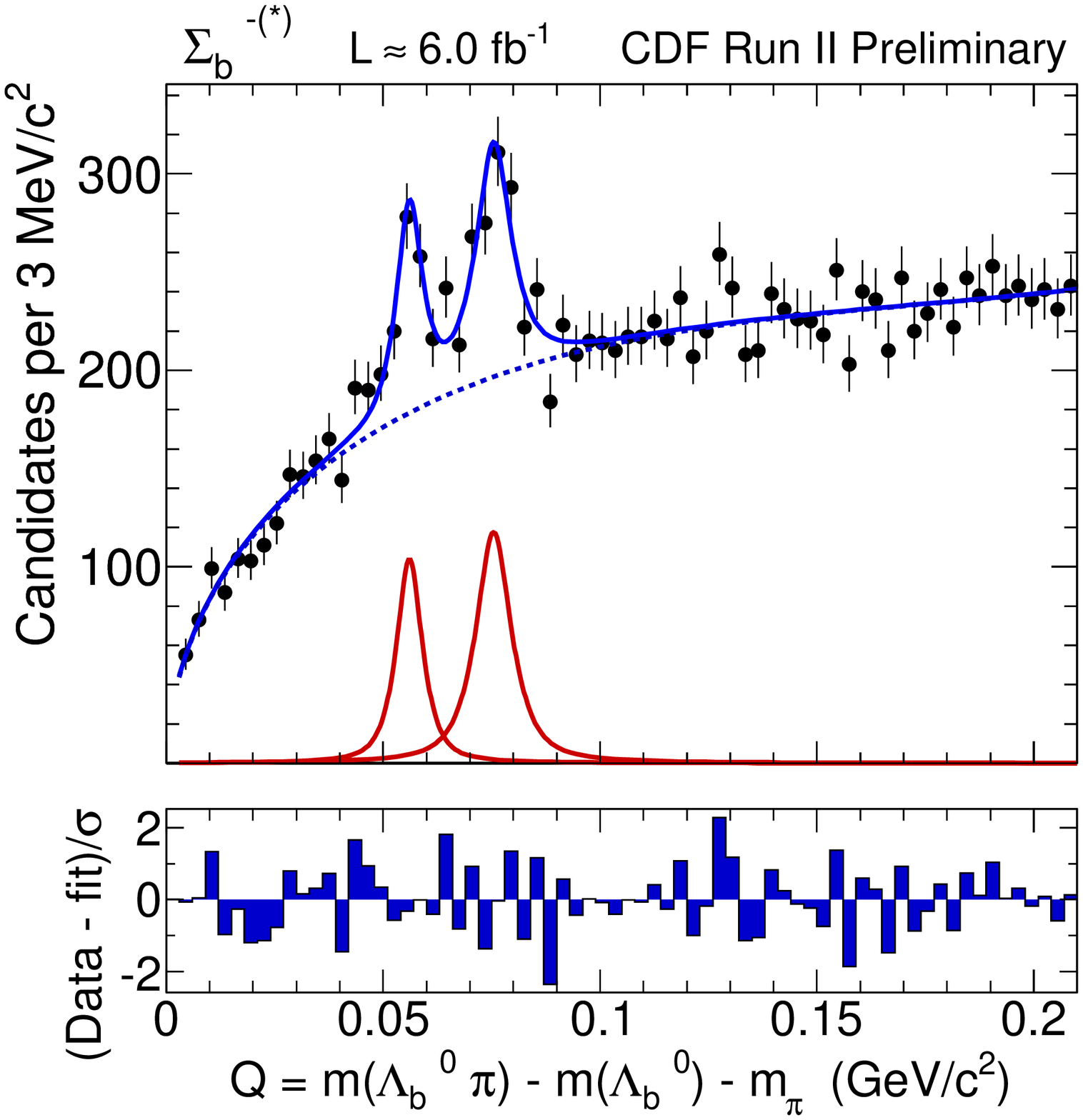}
\includegraphics[width=0.49\textwidth]{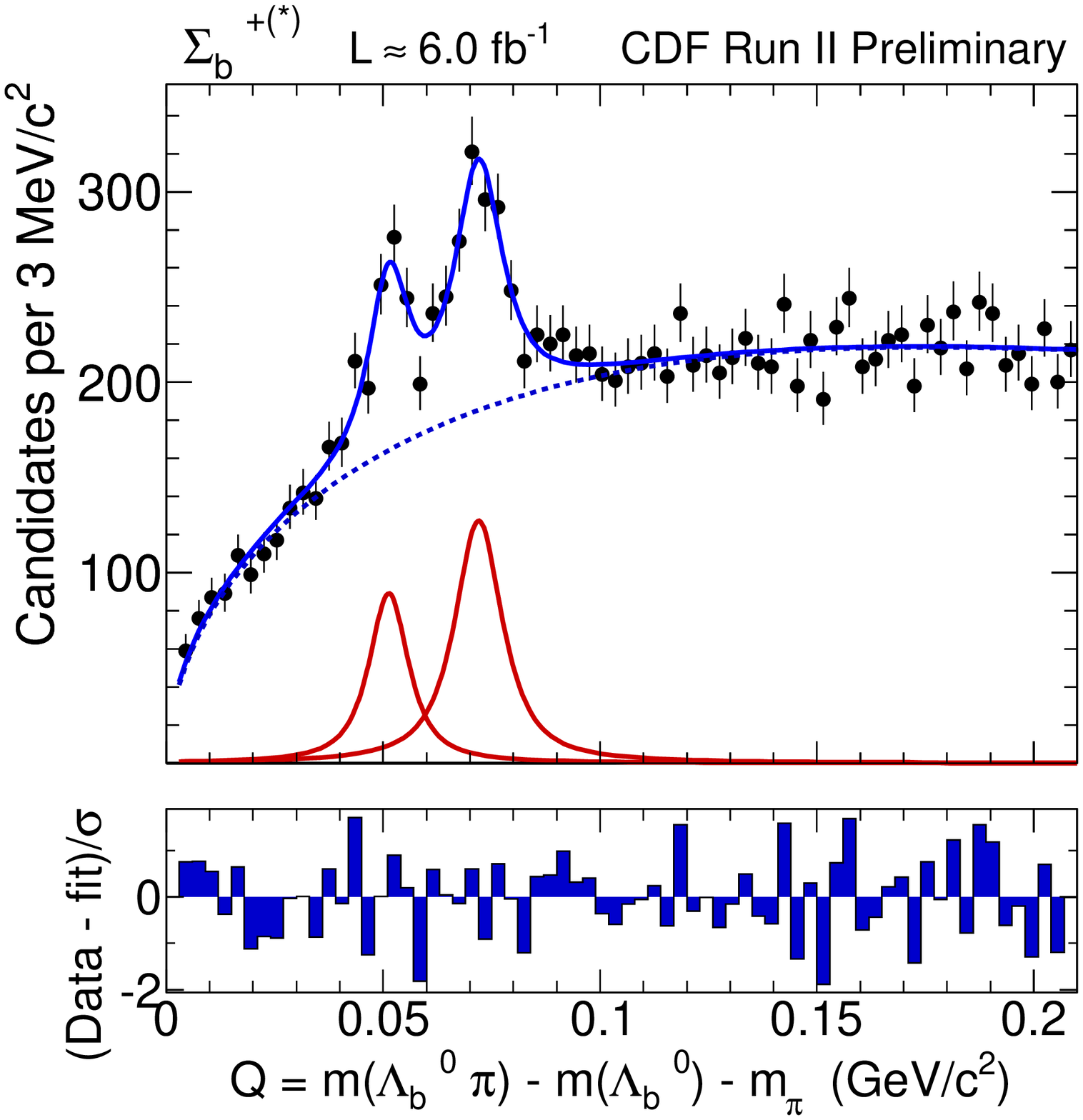}
\caption{$Q$ value distribution of \Sbsp (left) and \Sbsm (right) candidates.}
\label{fig:Sigmab}
\end{center}
\end{figure}

The hypotheses of having two, one, or no signal peaks are compared using a likelihood ratio.
For both isospin states the hypothesis of two peaks is favored by more than 7$\sigma$ over any of the other hypotheses, meaning that each peak has a significance above 7$\sigma$.

The measured masses and width are quoted in Tab.~\ref{tab:Sigmab}.
The dominant systematic uncertainty for the $Q$ values is the uncertainty on the momentum scale which is estimated from the comparison of reconstructed $\Sigma_c^{++}$, $\Sigma_c^0$, $\Lambda_c^{*+}$, and $D^{*+}$ masses with world average values.
The systematic uncertainty of the widths is dominated by the uncertainty on the resolution model estimated from $D^{*+} \rightarrow D^0\pi^+$ decays.
Further considered sources of systematic uncertainties are the background model and a fit bias.
When absolute masses are calculated from the $Q$ values the systematic uncertainty is limited by the knowledge of the \Lb mass.

\begin{table}[tb]
\begin{center}
\begin{tabular}{lcccc}
\hline
State & $Q$ value [MeV$/c^2$] & Mass [MeV$/c^2$] & Width [MeV$/c^2$] & Yield \\
\hline
\rule[-2mm]{0mm}{6mm} \Sbp  & $52.0^{+0.9}_{-0.8}\,^{+0.09}_{-0.4}$ & $5811.2^{+0.9}_{-0.8}\pm 1.7$ & 
$9.2^{+3.8}_{-2.9}\,^{+1.0}_{-1.1}$ & $468^{+110}_{-95}\,^{+18}_{-15}$ \\
\rule[-2mm]{0mm}{6mm} \Sbm  & $56.2^{+0.6}_{-0.5}\,^{+0.07}_{-0.4}$ & $5815.5^{+0.6}_{-0.5}\pm 1.7$ & 
$4.3^{+3.1}_{-2.1}\,^{+1.0}_{-1.1}$ & $333^{+93}_{-73}\pm 35$ \\
\rule[-2mm]{0mm}{6mm} \SbSp & $72.7\pm 0.7^{+0.12}_{-0.6}$ & $5832.0\pm 0.7\pm 1.8$ & 
$10.4^{+2.7}_{-2.2}\,^{+0.8}_{-1.2}$ & $782^{+114}_{-103}\,^{+25}_{-27}$ \\
\rule[-2mm]{0mm}{6mm} \SbSm & $75.7\pm 0.6^{+0.08}_{-0.6}$ & $5835.0\pm 0.6\pm 1.8$ & 
$6.4^{+2.2}_{-1.8}\,^{+0.7}_{-1.1}$ & $522^{+85}_{-76}\pm 29$ \\
\hline
\end{tabular}
\\[3mm]

\begin{tabular}{lc}
\hline
& Isospin splitting [MeV$/c^2$] \\
\hline
\rule[-2mm]{0mm}{6mm} $m(\Sbp) - m(\Sbm)$ & $-4.2^{+1.1}_{-0.9}\,^{+0.07}_{-0.09}$ \\
\rule[-2mm]{0mm}{6mm} $m(\SbSp) - m(\SbSm)$ & $-3.0\pm 0.9^{+0.12}_{-0.13}$ \\
\hline
\end{tabular}
\caption{Measurements of \Sbs properties. The first uncertainties are statistical, the second uncertainties systematic.}
\label{tab:Sigmab}
\end{center}
\end{table}

Compared with the previous analysis, the mass measurements are improved in precision by at least a factor two.
The isospin splittings and the natural widths are measured for the first time.

\section{\LbJpsiL Branching Ratio}

The quark level transition $b \rightarrow s$ is a flavor changing neutral current process and thus forbidden at tree level in the standard model and therefore considered a sensitive probe for new physics.
While such processes are well studied for $B$ mesons, little is known about $b \rightarrow s$ transitions in baryons.
The decay \LbJpsiL is kinematically very similar to the flavor changing neutral current decay $\Lambda_b \rightarrow \mu^+\mu^- \Lambda$ and was first observed by CDF in Run I~\cite{Abe:1996tr}.
However, the measured value of the \Lb production fraction times branching ratio, $f(b \rightarrow \Lb)\mathcal{B}(\LbJpsiL) = (4.7 \pm 2.3\mbox{ (stat.)} \pm 0.2\mbox{ (syst.)}) \times 10^{-5}$ has large uncertainties.

The D0 experiment presented a new measurement of this quantity using a data sample of 6.1~fb$^{-1}$ selected by a dimuon trigger~\cite{Abazov:2011wt}.
The $\Lambda$ and \Lb daughter particles are fitted to a vertex, respectively, and requirements on momenta, impact parameters and decay times are imposed that are optimized on the signal significance as estimated from simulation and data sidebands.
Cascade decays like $\Sigma \rightarrow \Lambda\gamma$ or $\Xi^0 \rightarrow \Lambda\pi^0$ are suppressed by requiring that the direction from the primary to the $\Lambda$ decay vertex coincides with the $\Lambda$ momentum direction.

Kinematically very similar $B^0 \rightarrow \Jpsi K^0_S$ decays with $K^0_S \rightarrow \pi^+\pi^-$ are used as normalization channel.
Figure~\ref{fig:LambdabBR} shows the invariant mass distributions of \Lb and $B^0$ candidates with a fit of a double Gaussian for signal and a second order polynomial for background.

\begin{figure}[htb]
\begin{center}
\includegraphics[width=0.49\textwidth]{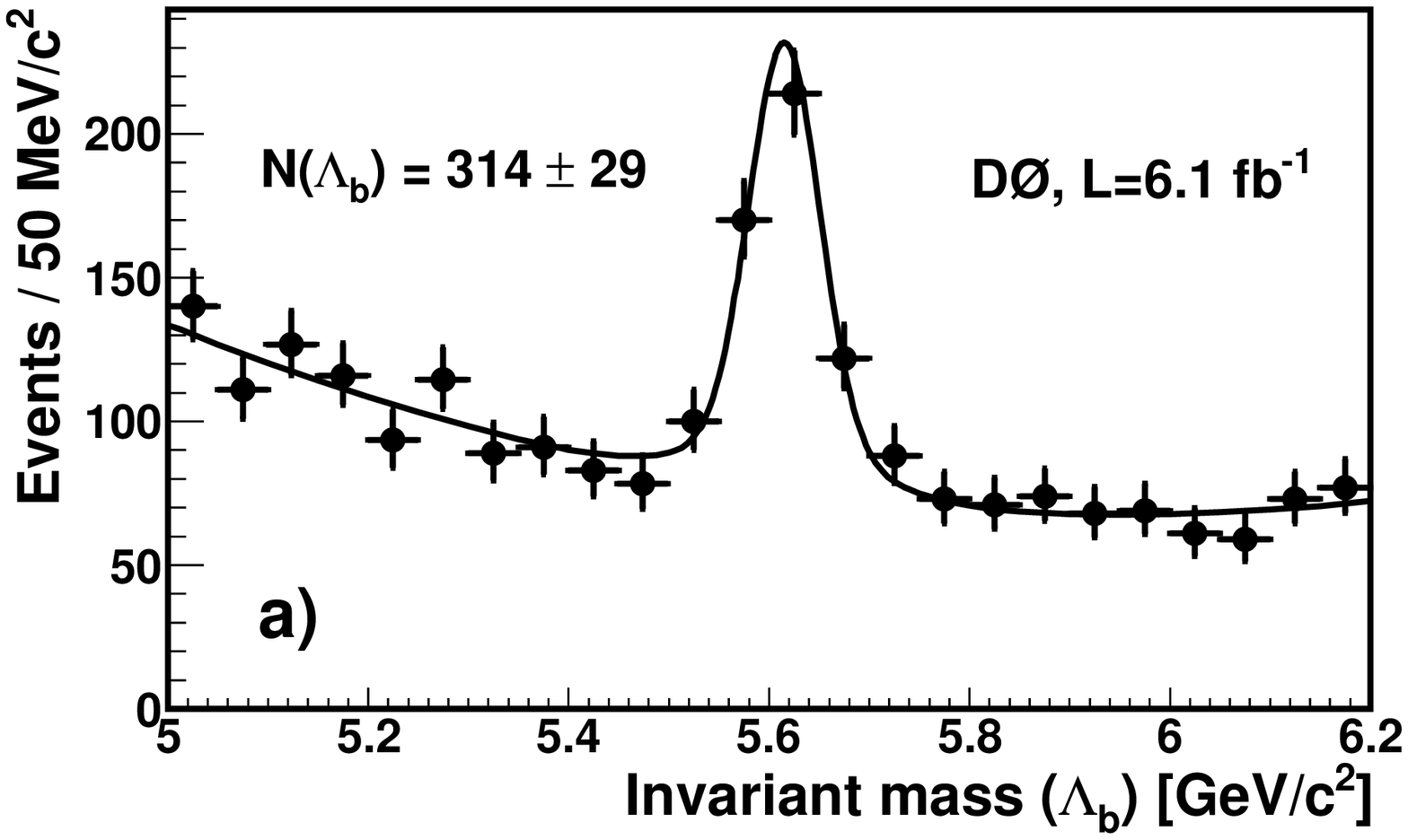}
\includegraphics[width=0.49\textwidth]{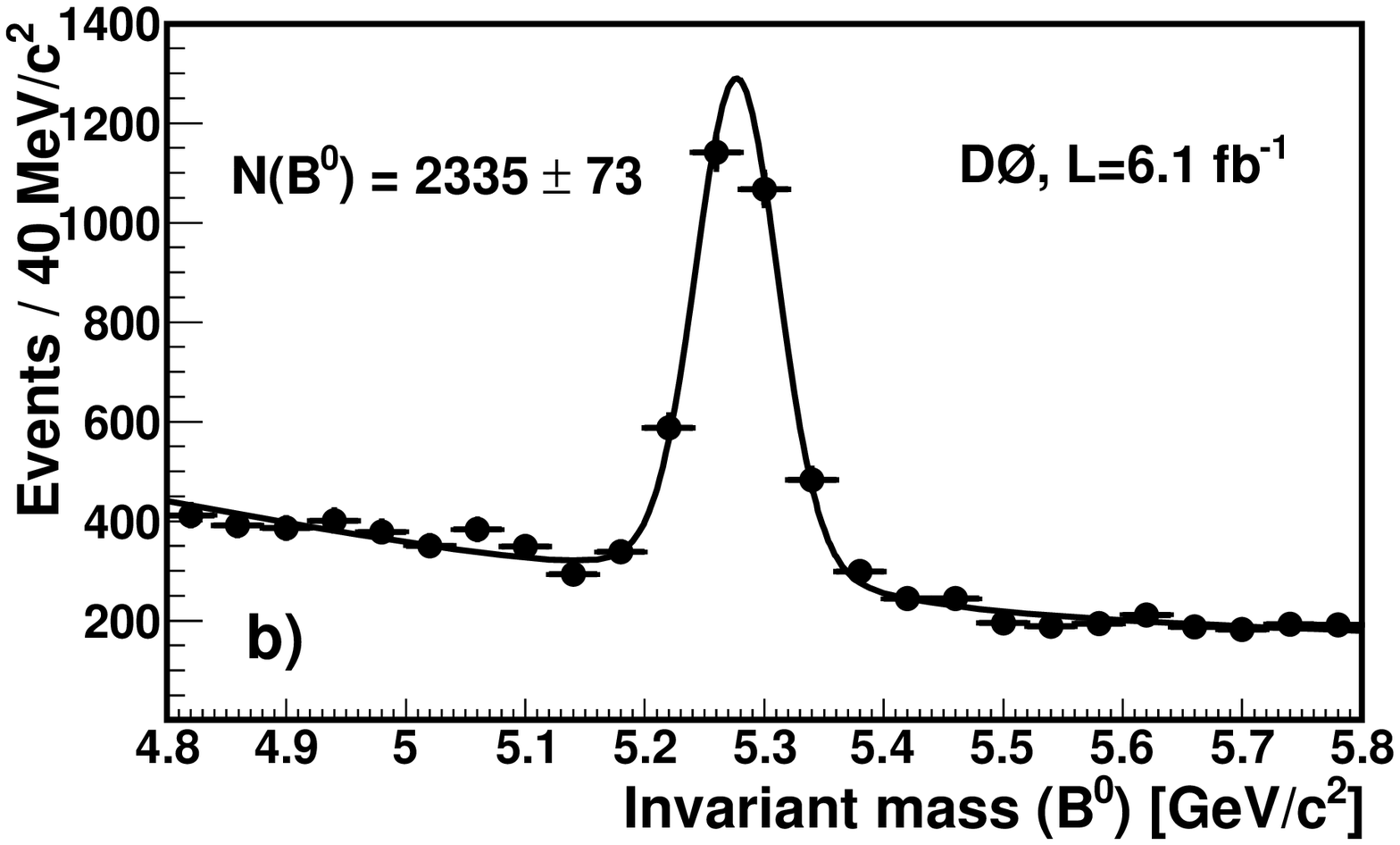}
\caption{Invariant mass distributions of \Lb (left) and $B^0$ (right) candidates.}
\label{fig:LambdabBR}
\end{center}
\end{figure}

Taking the relative efficiency from simulation, the following relative cross section is measured:
\[
\sigma_{rel} = \frac{f(b \rightarrow \Lb)\mathcal{B}(\LbJpsiL)}{f(b \rightarrow B^0)\mathcal{B}(B^0 \rightarrow \Jpsi K^0_S)} = 0.345 \pm 0.034\mbox{ (stat.)}\pm 0.033\mbox{ (syst.)} \pm 0.003 \mbox{ (PDG)}.
\]
The dominant systematic uncertainty comes from the unknown \Lb polarization, followed by the uncertainty due to the fit model.
The uncertainties on the relative efficiency and the cross-feed fractions have no significant influence.
Several cross checks were performed to verify the stability of the result.
Using the world average values for the $B^0$ production and decay fractions, a result of $f(b \rightarrow \Lb)\mathcal{B}(\LbJpsiL) = (6.01 \pm 0.60\mbox{ (stat.)} \pm 0.58\mbox{ (syst.)} \pm 0.28\mbox{ (PDG)}) \times 10^{-5} = (6.01 \pm 0.88) \times 10^{-5}$ is obtained which is about three times more precise than the CDF Run I measurement.

\section{Charm Baryons}

The Tevatron experiments have not only provided significant contributions on the $b$ baryon sector, but have also collected large samples of charm baryon decays which allow studies at unprecedented precision.
The spectrum and decay modes of the \Scs and \LcS states studied by CDF~\cite{Aaltonen:2011sf} are the same as the ones shown in Fig.~\ref{fig:bBaryonSpectrum}, but for the charm instead of the $b$ baryon sector.

In a data sample of 5.2~fb$^{-1}$, $\Lc \rightarrow p K^- \pi^+$ decays are selected by a trigger on displaced tracks.
Since charm hadrons have a shorter lifetime than $b$ hadrons, about half of the triggered \Lc particles come from $b$ baryon decays.
\Lc candidates are combined with one or two pion tracks to form \Scs or \LcS candidates, respectively.
Neural networks are used for the selection of \Lc, \Scs, and \LcS candidates.
The networks are trained with data only using the $_sPlot$ technique.

\begin{figure}[htb]
\begin{center}
\includegraphics[width=0.49\textwidth]{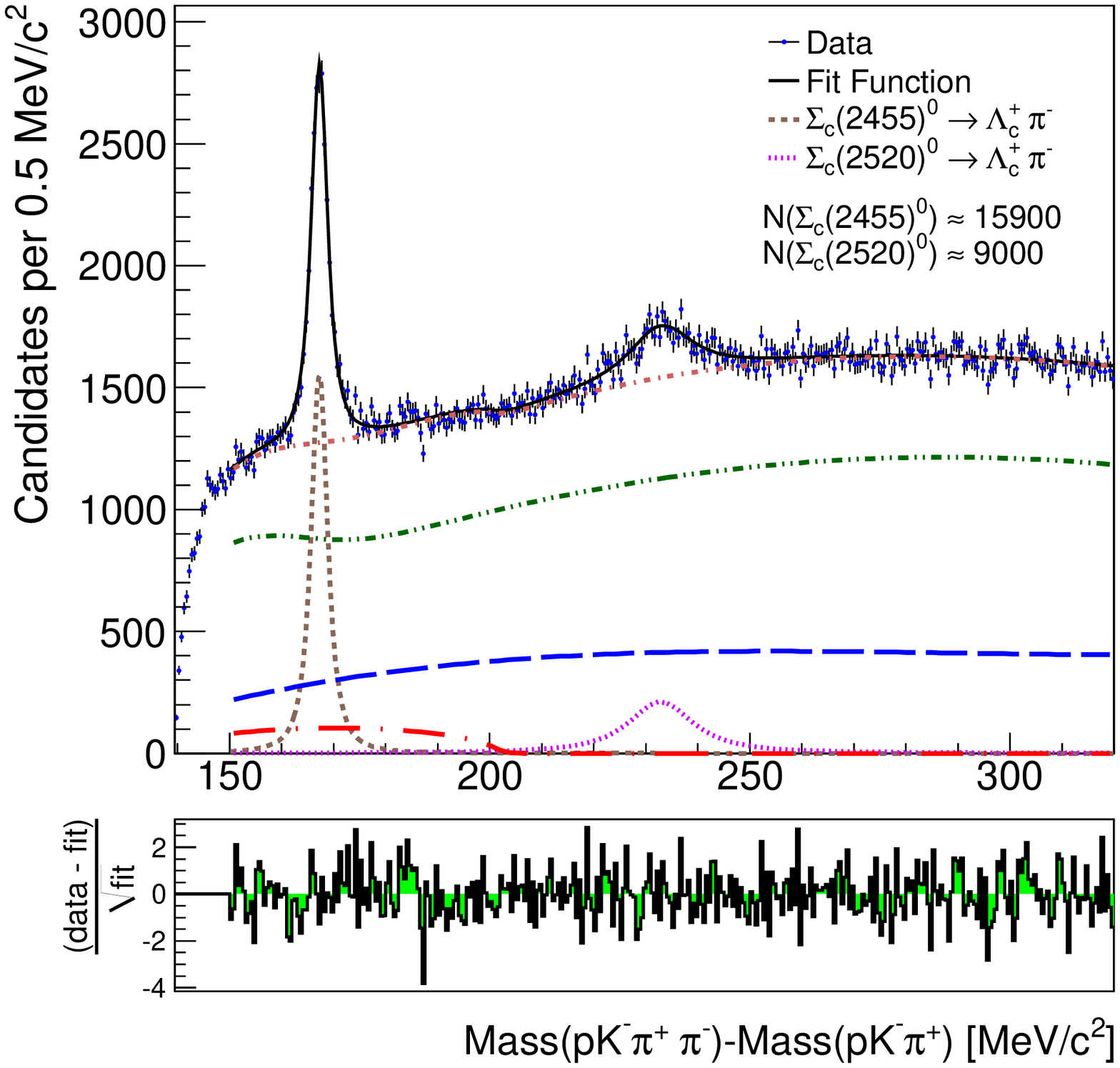}
\includegraphics[width=0.49\textwidth]{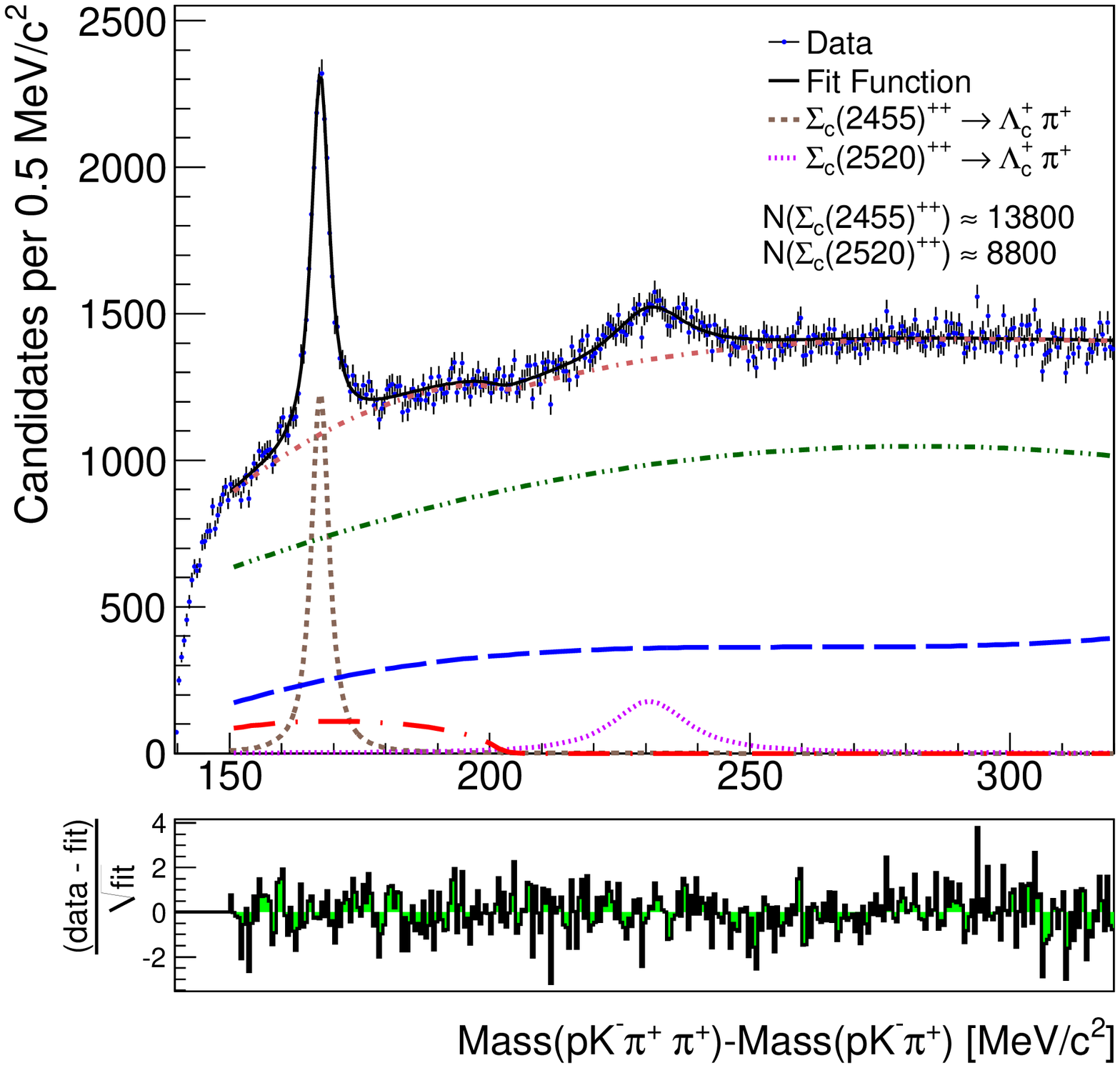}
\caption{Mass difference distributions of $\Sigma_c^{(*)0}$ (left) and $\Sigma_c^{(*)++}$ (right) candidates.}
\label{fig:Sigmac}
\end{center}
\end{figure}

\begin{figure}[htb]
\begin{center}
\parbox[c]{0.49\textwidth}{\includegraphics[width=0.48\textwidth]{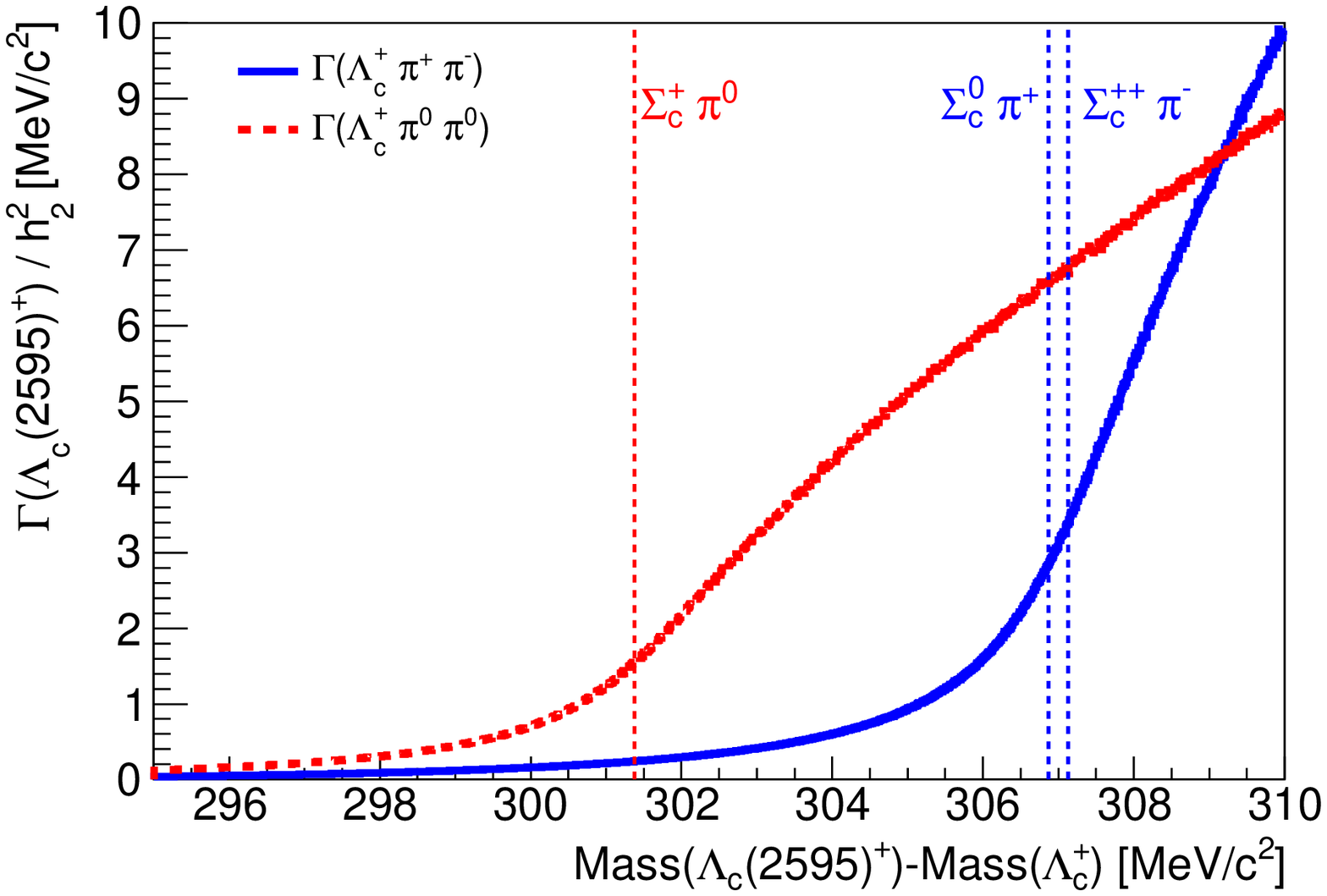}}
\parbox[c]{0.49\textwidth}{\includegraphics[width=0.48\textwidth]{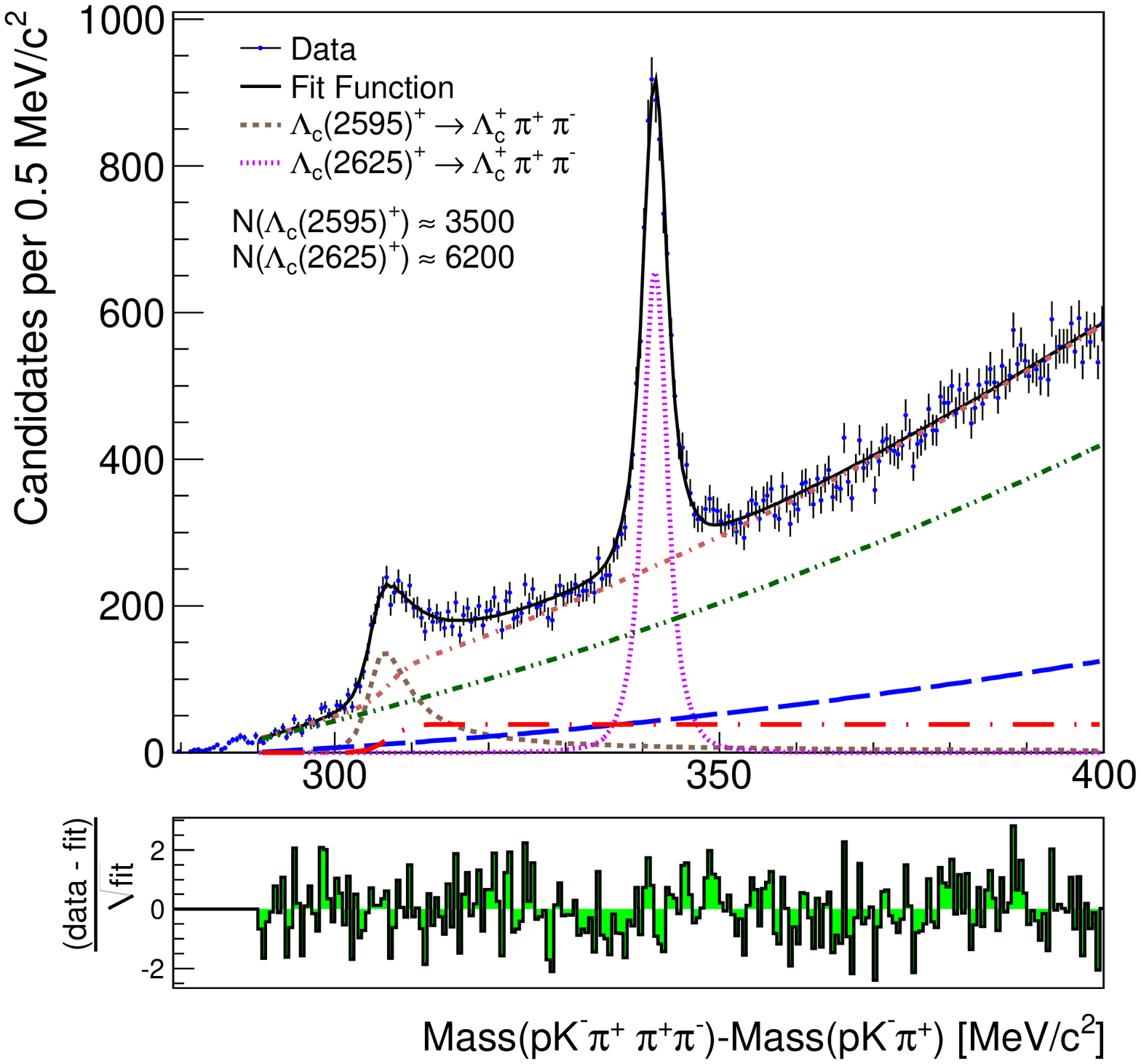}}
\caption{Mass dependent \Lcl partial decay widths (left) and mass difference distribution of \LcS candidates (right).}
\label{fig:Lambdac}
\end{center}
\end{figure}

Distributions of mass differences relative to the reconstructed \Lc mass are fitted to measure mean masses and widths.
The distributions for the \Scs states are shown in Fig.~\ref{fig:Sigmac}.
The signals are described by a nonrelativistic Breit-Wigner convolved with a triple Gaussian resolution function.
The combinatorial background (green dash-dot-dotted line) is described by a second order polynomial with parameters determined from \Lc sidebands.
In the case of the neutral \Sc states an additional component from $D^*$ reflections has to be taken into account which is parametrized by a Gaussian.
Real \Lc with a random pion (blue dashed line) are parametrized by a third order polynomial with all parameters free in the fit.
The third background component comes from $\Lc(2625) \rightarrow \Lc\pi^+\pi^-$ decays (red dash-dotted line).
It is determined using the $\Lc(2625)$ yield measured in data.
$\Lc(2595)$ particles decay mainly resonantly and thus contribute to the signal.

The fit to the mass difference distribution of \LcS candidates is shown in Fig.~\ref{fig:Lambdac} (right).
Signals and backgrounds are treated in the same way as in the fits to the \Scs distributions, except that the cross-feed background now comes from \Scs decays and the threshold effect of $\Lcl \rightarrow \Sc \pi$ has to be taken into account.
This is done by using a mass dependent width as illustrated in Fig.~\ref{fig:Lambdac} (left).
The parameter determining the width of the lineshape is the pion coupling constant $h_2$.
CDF has shown that a mass independent width, as used in previous analyses, does not describe the data.
Because of the low signal yield in previous experiments of up to $112\pm17$ events~\cite{Edwards:1994ar}, the threshold effect was not observable so far.
As a consequence of the threshold effect, the mean \Lcl mass is significantly shifted towards lower values.

\begin{table}[tb]
\begin{center}
\begin{tabular}{ccc}
\hline
State & $\Delta m$ [MeV$/c^2$] & $\Gamma$ [MeV$/c^2$] \\
\hline
$\Sigma_c(2455)^{++}$ & $167.44 \pm 0.04 \pm 0.12$ & $2.34  \pm 0.13 \pm 0.45$\\
$\Sigma_c(2455)^{0}$  & $167.28 \pm 0.03 \pm 0.12$ & $1.65  \pm 0.11 \pm 0.49$\\
$\Sigma_c(2520)^{++}$ & $230.73 \pm 0.56 \pm 0.16$ & $15.03 \pm 2.12 \pm 1.36$\\
$\Sigma_c(2520)^{0}$  & $232.88 \pm 0.43 \pm 0.16$ & $12.51 \pm 1.82 \pm 1.37$\\
$\Lambda_c(2595)^{+}$ & $305.79 \pm 0.14 \pm 0.20$ & $h_2^2 = 0.36 \pm 0.04 \pm 0.07$\\
$\Lambda_c(2625)^{+}$ & $341.65 \pm 0.04 \pm 0.12$ & $< 0.97$ at 90\% CL\\
\hline
\end{tabular}
\label{tab:FitResults}
\caption{Charm baryon properties. The first uncertainties are statistical and the second systematic.}
\label{tab:CharmBaryons}
\end{center}
\end{table}

\begin{figure}[htb]
\begin{center}
\includegraphics[width=0.49\textwidth]{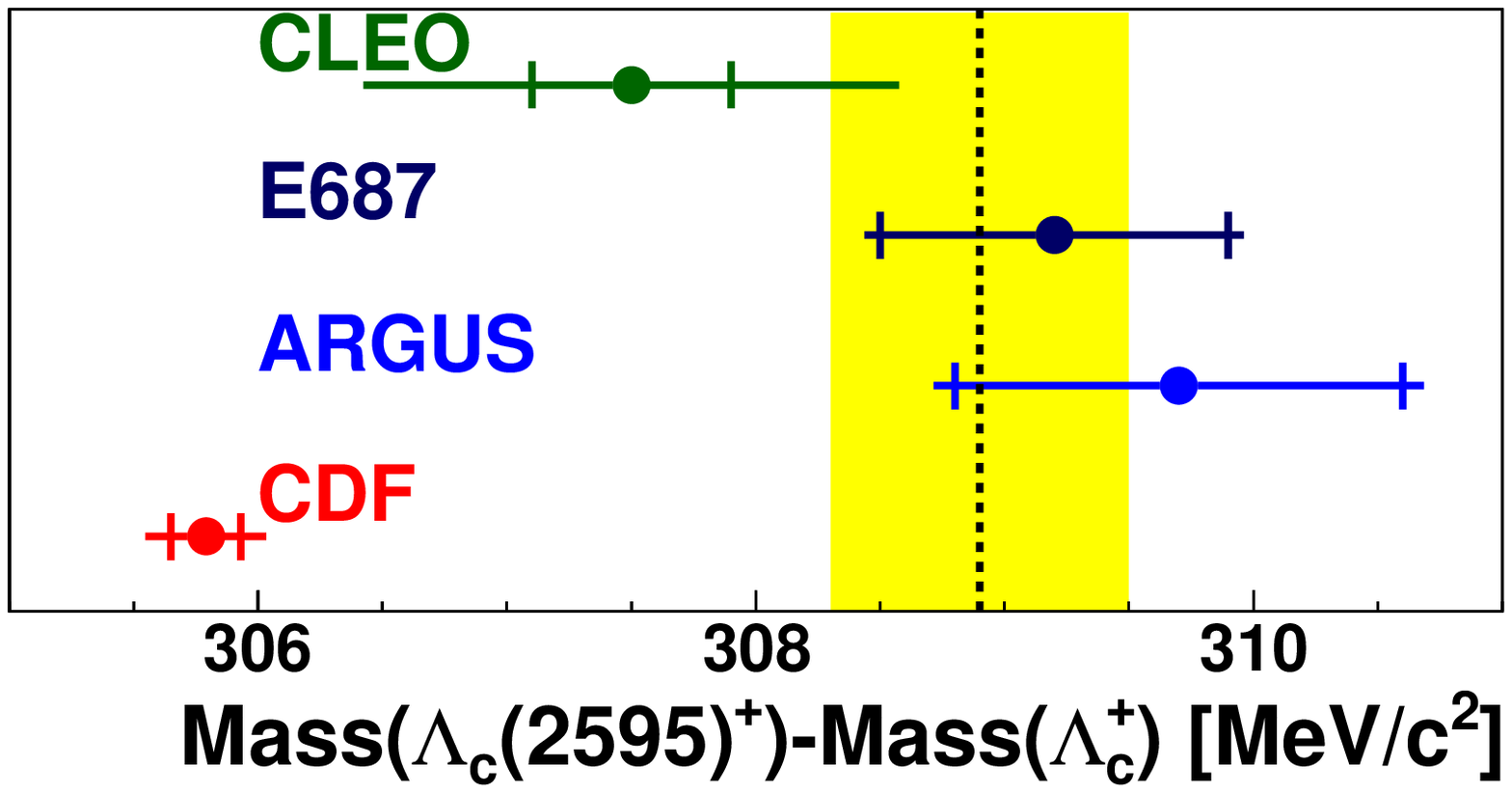}
\includegraphics[width=0.49\textwidth]{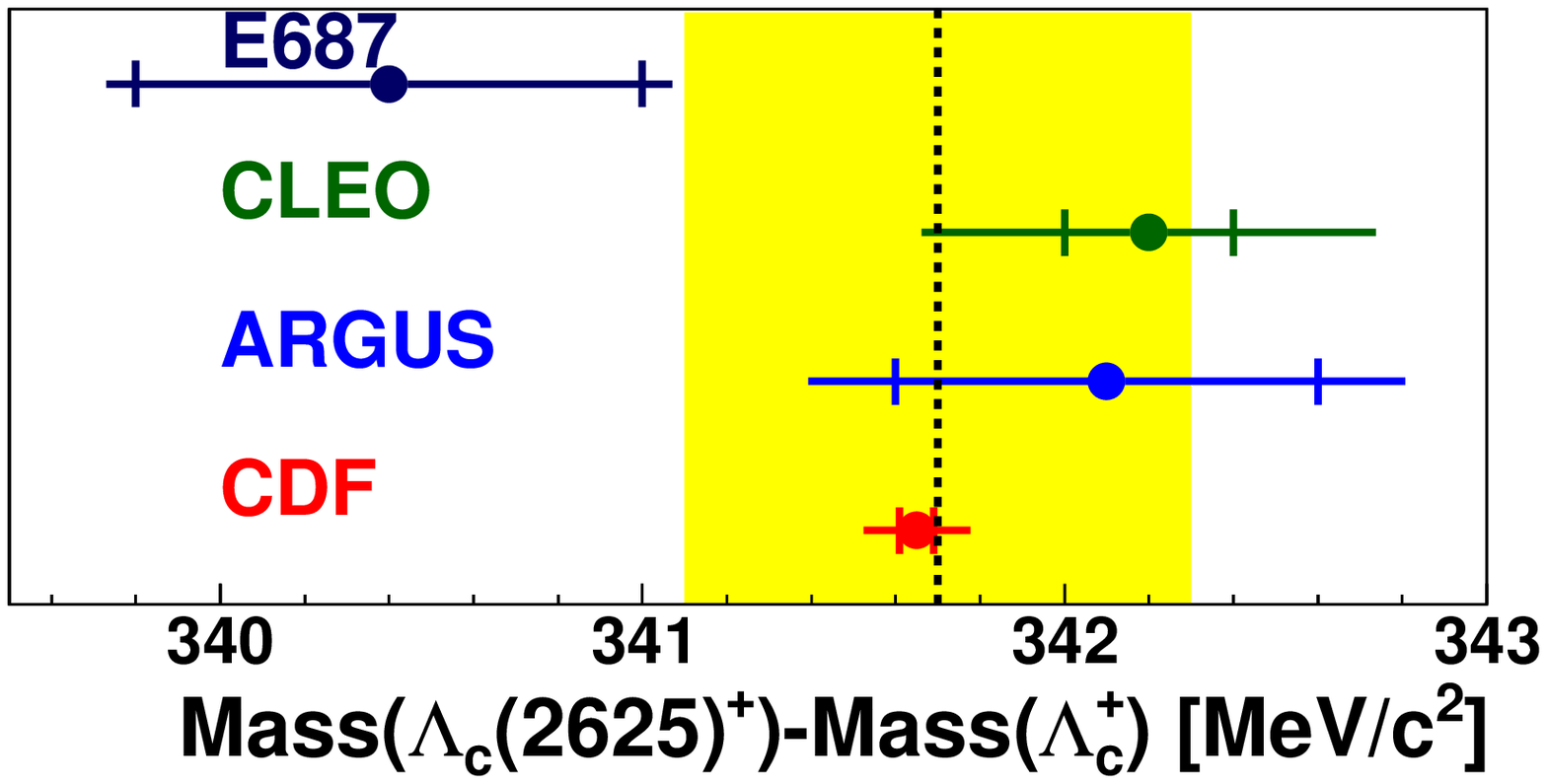}
\caption{Comparison of \LcS mass difference measurements by CLEO ~\cite{Edwards:1994ar}, Fermilab E687~\cite{Frabetti:1993hg,Frabetti:1995sb}, ARGUS~\cite{Albrecht:1997qa}, CDF and the world average without the CDF result (yellow band).}
\label{fig:LambdacComparison}
\end{center}
\end{figure}

Numerical results of the measured masses and widths are given in Tab.~\ref{tab:CharmBaryons}.
The \Scs results agree well with the world average values and are of comparable precision.
A significant improvement in precision of the \LcS propertied and observation of the threshold effect in the \Lcl line shape as predicted in Ref.~\cite{Blechman:2003mq} are achieved as illustrated in Fig.~\ref{fig:LambdacComparison}.

\section{Conclusions}

So far our knowledge about baryons containing a $b$ quark fits on one page in the PDG particle list table~\cite{Nakamura:2010zzi}.
Thus the observation of new states and improved or even first measurements of heavy baryon properties are very welcome to improve the phenomenological models of strongly bound states and to advance on the way to a deeper understanding of low energy QCD.
The Tevatron experiments CDF and D0 have been very active in the last few years and contributed significantly to our knowledge of heavy baryons.
With the good performance of the Tevatron both experiments could continue to present new and updated results on this sector as shown in this article.



%

}  



\begin{thebibliography}{99}
  
\bibitem{Aaltonen:2007rw}
  T.~Aaltonen {\it et al.} (CDF Collaboration),
  Phys.\ Rev.\ Lett.\  {\bf 99}, 202001 (2007).

\bibitem{Abazov:2007ub}
  V.~M.~Abazov {\it et al.}  (D0 Collaboration),
  Phys.\ Rev.\ Lett.\  {\bf 99}, 052001 (2007).

\bibitem{Aaltonen:2007un}
  T.~Aaltonen {\it et al.}  (CDF Collaboration),
  Phys.\ Rev.\ Lett.\  {\bf 99}, 052002 (2007).

\bibitem{Buskulic:1996sm}
  D.~Buskulic {\it et al.} (ALEPH Collaboration),
  Phys.\ Lett.\  B {\bf 384}, 449 (1996).

\bibitem{Abreu:1995kt}
  P.~Abreu {\it et al.}  (DELPHI Collaboration),
  Z.\ Phys.\  C {\bf 68}, 541 (1995).

\bibitem{Abazov:2008qm}
  V.~M.~Abazov {\it et al.} (D0 Collaboration),
  Phys.\ Rev.\ Lett.\  {\bf 101}, 232002 (2008).

\bibitem{Aaltonen:2009ny}
  T.~Aaltonen {\it et al.} (CDF Collaboration),
  Phys.\ Rev.\  D {\bf 80}, 072003 (2009).

\bibitem{Triggers}
  L.~Ristori, G.~Punz,
  Annu.\ Rev.\ Nucl.\ Part.\ Sci.\ \textbf{60}, 595 (2010).

\bibitem{Aaltonen:2011wd}
  T.~Aaltonen {\it et al.} (CDF Collaboration),
  arXiv:1107.4015 [hep-ex].

\bibitem{cdf10286}
  CDF Collaboration,
  CDF note 10286.

\bibitem{Abe:1996tr}
  F.~Abe {\it et al.} (CDF Collaboration),
  Phys.\ Rev.\  {\bf D 55}, 1142-1152 (1997).

\bibitem{Abazov:2011wt}
  V.~M.~Abazov {\it et al.} (D0 Collaboration),
  Phys.\ Rev.\  {\bf D 84}, 031102 (2011).

\bibitem{Aaltonen:2011sf}
  T.~Aaltonen {\it et al.} (CDF Collaboration),
  Phys.\ Rev.\  {\bf D 84}, 012003 (2011).

\bibitem{Edwards:1994ar}
  K.~W.~Edwards {\it et al.} (CLEO Collaboration),
  Phys.\ Rev.\ Lett.\  {\bf 74}, 3331-3335 (1995).
  
\bibitem{Blechman:2003mq}
  A.~E.~Blechman, A.~F.~Falk, D.~Pirjol, J.~M.~Yelton,
  Phys.\ Rev.\  {\bf D 67}, 074033 (2003).

\bibitem{Frabetti:1993hg}
  P.~L.~Frabetti {\it et al.} (E687 Collaboration),
  Phys.\ Rev.\ Lett.\  {\bf 72}, 961-964 (1994).

\bibitem{Frabetti:1995sb}
  P.~L.~Frabetti {\it et al.} (E687 Collaboration),
  Phys.\ Lett.\  {\bf B365}, 461-469 (1996).
  
\bibitem{Albrecht:1997qa}
  H.~Albrecht {\it et al.} (ARGUS Collaboration),
  Phys.\ Lett.\  {\bf B402}, 207-212 (1997).
  
\bibitem{Nakamura:2010zzi}
  K.~Nakamura {\it et al.} (Particle Data Group Collaboration),
  J.\ Phys.\ G {\bf 37}, 075021 (2010).
  

\end{thebibliography}
\end{document}